**Universal Service in times of Reform:**

**Affordability and accessibility of telecommunication services in Latin America**

**Martha Fuentes-Bautista**

**(student)**

**University of Texas at Austin**

**Department of Radio-TV-Film**

**CMA 6.118**

**Austin, TX 78712**

**(512) 478-8870**

*Title*

*Universal Service in times of Reform: affordability and accessibility of telecommunication services in Latin America*

*Abstract*

By surveying the universal service policies in six Latin American countries, this study explores the evolution of the concept during the rollout of the telecommunication reform in the last decade. Country profiles and a set of universal service indicators provide a frame for discussing issues of accessibility and affordability of telephone service in the region. This study found that the reconfiguration of national networks fostered by liberalization policies offered risks and opportunities to achieve universal service goals. The diversification of access points and services enhanced users choices but price rebalancing and lack of Universal Service Obligations (USO) to target groups with special needs depressed the demand and threatened to exclude significant parts of the population. The situation requires the reformulation of USO incorporating all technological solutions existing in the market, and factors from the consumer-demand accounting for the urban-rural continuum, and different social and economic strata. This study identifies the emergence of a second generation of USO targeting some of these needs. However, more competition and special tariff plans for the poor need to be incorporated to the options available in the market.

**INTRODUCTION**

The term *universal service*, as first used in the AT&T slogan "one system, one policy, universal service," was part of a corporate strategy aimed at undermining the position of multiple rival networks that served the United States in 1907. The AT&T strategy was to peddle the idea of consolidating independent telephone exchanges into local monopolies that could interconnect as many users as possible (Mueller, 1997). The benefits of a pervasive network that overcame the rather fragmented and disconnected scenario of previous years was an excellent showcase for this model. One result was that the achievement of universal service was subordinated to the existence of a monopoly for basic telephony for decades. The social dimension of the concept grew up along with spawning public telephone and telegraph monopolies (PTTs) that served countries in Europe, Africa, Asia and Latin America for more than 80 years. However, with the exception of the wealthiest nations, the rate of investment of PTTs was insufficient to meet the demand of telecommunication services in most nations.

Since the mid 1980s liberalization policies seeking to achieve industry efficiencies have emerged as an alternative path to attain the goal of universal service. These waves of liberalization struck Latin America with particular energy. The International Telecommunication Union (ITU) (2000) estimates that, between 1988 and 1998, more than two-thirds of the Latin American PTTs were partially or fully privatized. Animated by capital that by and large came from wealthier nations, basic telephone networks in the region have grown twice as fast as the telephone systems of developed countries. Despite this burst of capital and services, not much more than one-third of the region's households have a fixed phone (ITU, 2000:3). Achieving 100% service penetration is still a pending goal in developing nations that have pursued liberalization policies. The issue becomes even more urgent to address in light of the rapid growth of the Internet or computer-mediated communications (CMC) in developed nations, a situation that threatens to widen the gap between the First and the Third world.

By reviewing the cases of Argentina, Brazil, Chile, Mexico, Peru and Venezuela, this study traces the evolution of the universal service policies designed and pursued during the first decade of liberal reforms. This study addresses broader questions of how different approaches to telecommunication reform are fostering higher levels of service penetration in developing countries, and what types of governance they have brought about. The goal is to identify particular trends of telecommunication services penetration, and to draw lessons that help to improve the strategy of network development of these nations, most of which are currently engaged in a second round of legislative reforms.

# THEORETICAL FRAMEWORK

## <u>Universal service: a concept in transformation</u>

The basic conceptualization of universal service in the telecom business is commonly associated with the availability of telephone lines accessible to the public, in both economic and geographical terms. As Mueller puts it, it is "a telephone network that covers all of a country, is technologically integrated, and connects as many citizens as possible" (1997, p.1). This ambitious parameter calls for no distinctions between rural and urban areas, economic strata or social categories.

Universal service is pursued through diverse means that usually focus on the supply-side of telecommunication services. Universal Service Obligations' (USO) strategies include:

(1) Extending the service to unserved populations in inner city and rural areas.

(2) Meeting unsatisfied demand for access lines targeting customers on waiting lists.

(3) Making telecommunication services affordable for low-income citizens, commonly known as the 'unphoned'. Strategies aiming at this goal go from the expansion of pay phone networks (public telephony) to tariff systems that benefit the poorest households. (Tyler, Letwin & Roe, 1995).

Focusing on infrastructure accessibility, or service affordability and reliability, universal service policies are key to assure network expansion beyond the limits established by free market dynamics. Despite the apparently clear definition and policy objectives, a technical description of what is "essential" in telecommunications has proven to be a complex and controversial topic. Technological convergence and blurring market barriers have altered the structure of the telecommunication business as well as traditional definitions of *basic service*. Teledensity or the number of fixed phone lines per hundred inhabitants, the index traditionally used to weigh the concept, is nowadays considered an "imperfect measure of universal service" (ITU, 1998b, p.20). Some have proposed instead to use household service penetration as an indicator to gauge universal service (Mueller, 1996; ITU, 1998b)[1], since it indicates residential use. However, neither of these indices reflects the direction and segmentation of a network's expansion, nor do they account for other access methods such as mobile telephony, satellite and digital channels, which are currently used by citizens of developing and developed countries to access national and global networks.

---

1 The strength of this indicator is based on three aspects: a) a phone at home is considered the closest connection of citizens in times of emergency or distress; b) it captures countries' specificities such as different household sizes; c) and it is related to income level. The ITU (1998b) considers that countries with household penetration of 90% or above to have reached universal service.

The rapid pace of technological developments, and economic constraints still faced by developing markets are also inducing administrators to review the notion of *universality* in two directions. The first, concerned with new telecommunication services *availability*, suggests an enhanced version of the network that encompasses main telephone lines, mobile cellular subscribers and digital access. The second focuses on *accessibility* to the network and stresses the existence of an access point (e.g. a telephone) within reasonable distance for everyone. Table 1 summarizes some indicators related to each concept.

**Table 1. Universal Service and Universal Access Indicators**

| Universal Service (availability) | Universal Access (accessibility) |
|---|---|
| % of households with a telephone | Access to a telephone established in terms of: |
| Mobidensity (Number of fixed lines plus mobile cellular lines per hundred inhabitants) | - time or distance from a telephone<br>- minimum population units per phone |
| Digital access (ISDN channels) | Payphones per 1000 inhabitants/ or per main lines |

Source: ITU (1998b)

Some authors argue that these concepts address different problems and are not mutually exclusive. Service availability and accessibility can complement each other to the degree to which new technologies capable of supporting information services (wireless, satellite) increase the means of interconnection to national networks, making them more affordable to the public. Hudson (2000) considers that the economic and demographic diversity of developing countries demand a "*multi-leveled*" definition of access identifying expansion targets within communities, institutions and households. Minges (1999a, 1999b) advocates for a similar model that places *universal access* as a step towards *universal service*. He illustrates the situation through the example of the recent evolution of the Swiss telephone system. Between 1995 and 1998, the teledensity in Switzerland decreased from 60.6 to 56 main lines/100 people, as a result of residential and commercial customers migrating to mobile cellular services and ISDN networks. However, the author argues, the network as an aggregate -including all three services- actually grew more than 20%, adding up to a million subscribers over the same period.

Responding to completely different realities and needs, the mobile cellular networks of countries such as in Lesotho, Cote d'Ivoire, Paraguay and Venezuela have reached penetrations equivalent to half of the national teledensity (ITU 1999, 2000). Wireless communications are used to overcome geographic and financial barriers commonly encountered in the deployment of services in developing nations. Although the experience there is not new, it does indicate how technologies that act as product substitutes for the fixed telephone can enhance our definition of basic services[2].

Attending to the changing meaning of *basic service* and *universality*, Hudson proposes that developing countries should understand universal access as a "dynamic concept with a set of moving targets".  It should not tie goals to specific technology or service provider (e.g. wired or wireless service), but rather state them in terms of "functions and capabilities, such as ability to transmit voice and data" (2000:3). This view can open to administrators new alternatives for policy design through a "*multi-network*" perspective, one that considers the total sum of capabilities enabling a citizen's access to a national telecommunication network. Criticism of the "*multi-network*" perspective arises from the difficulties in accounting for penetration of new services (e.g. mobile telephony) by social strata and geographical area. The ITU considers that broadening the definition to these services "may lead to telecommunications investment becoming narrowly focused upon those who can afford to pay for the new services thereby increasing regional disparities and exacerbating information poverty." (1998b, p.85).

In its latest review of the concept, and despite important transformations in technology and the structure of markets, the ITU has assumed a fairly traditional definition of universal service, avoiding major discussions about possible redistribution mechanisms for telecommunications investment, and new statistical mechanisms allowing higher accountability. In the 1998 World Telecommunication Development Report, the organization concludes that there is  "no compelling reason, at present, to expand the definition of universal service to include individual access to information services" (85). The report advocates for a "practical definition of universal access" that targets "increased levels of household telephone penetration" in developed countries, and "community access through the provision of public telephones" in developing nations (p.90).

This conceptual divergence is interpreted as the result of the growing gap between the poorest nations that are struggling to build a basic telephone network, and the wealthiest countries already engaged in the creation of the information superhighway. As Gil-Egui and Steward assert, the consequence of such a dichotomy is a "dual rhetoric that speaks of universality of plain old telephone for the Third World and broadband for the First World" disregarding the higher upgrading cost that the poorest nations will have to face in the future (Gil-Egui & Steward, 1999, p. 2).

However, this paper argues that the inherent contradictions of such distinctions can be overcome through a multi-network approach that extends USO to all technological solutions existing in the market and takes into account user' preferences and active role in defining different levels to gain access to the national network. This would require further development of competition as well as more effective and flexible regulation to target market imbalances in dynamic technological scenarios. In the dawn of a second generation of telecommunication reforms in developing countries, an assessment of universal service policies played out in these nations during a period better characterized as the transition to total competition can shed light on available paths to attain universal service.

---

[2] In fact, the WTO Secretariat recognizes the multi-access notion through the inclusion of mobile services and circuit-switched data transmissions, along with voice telephone services, in the list of services ruled by the 1997

### Risks and opportunities of liberalization

Since the onset of liberalization, a considerable amount of research has tried to test the implications of telecommunication reform attending to economic and technological criteria. As Mosco (1988) warns, few of these efforts have gone beyond dichotomist discussions that oppose public to private, and government control to free market. Although this research has produced detailed descriptions of the new industry structure, pricing strategies and regulations, most of the time it misses the opportunity to assess the way in which market constraints have been rebuilt, and the implications of such changes for diverse social groups. Different approaches to liberalization have included a diverse mix of ownership strategies (privatization, corporatization, strategic partnership, leasing of basic infrastructure), and levels of market competition (private monopoly, duopolies, competing product substitute markets).  This diversity of strategies makes it virtually impossible to talk about a typology of approaches to telecommunication reform. However, efforts should be made to assess the social implications of the different market structures and practices brought about by liberalization.

Here, I shall suggest that one way to discuss the different approaches to reform can be related to the reconceptualization of USO and other conditions for network expansion, which have particularly important repercussions for developing countries embarked on those reforms. Through USO, the State reformulates the relationship between operators, the public and the regulator. USO also targets market inefficiencies, setting some conditions for operation without hampering the economics of the business. They should be designed to meet the needs of particular populations with problems of access attending to claims of the public.

Comparative studies of telecommunications reform in developing countries have mainly focused on the dichotomy of controlled versus free markets. Typically, these analyses contrast the outcomes of liberalized versus regulated scenarios in an attempt to identify the benefits or setbacks of liberalization. A number of different studies have found in common an immediate "efficiency effect" after reform, with a tendency for slow-down overtime (Grande, 1994; Petrazzini, 1995; Straubhaar et

---

GATS commitments or limitations on market access and national treatment for basic telecommunications.

al., 1995). General assessments on the impact of liberalization policies on developing countries lend weight to a positive view that shows increasing penetration at the household and small firm levels (Petrazzini, 1996).

Other studies have explored the impact of privatization on network growth. In their account of teledensity evolution in Asian and Latin American countries that underwent liberalization policies in the first half of the 1990s, Petrazzini and Clark (1996) found that "in countries with privatized telecommunication (penetration) grew twice as fast compared with countries that did not privatize" (p.37). The authors also examine the effect of competition on the rollout of new services such as cellular phones. Testing variables possibly affecting cellular density growth such as GDP levels, researchers found that none of them was as important as competition.

More recently, Jayakar (September, 2000) empirically tests changes in residential teledensity in 30 countries as a function of economic growth, diffusion effects and three policy variables: ownership, industry structure and regulatory independence. The research identified diffusion effects and policy variables as significant predictors of telephone penetration at the household level. The author tests the model accounting for possible differences between rich and poorer countries. Interestingly, variables such as privatization or de-monopolization were found not significant as explanatory variables of growth in household service penetration. Gutierrez and Berg (2000) examine institutional, regulatory and economic determinants of telephone penetration in Latin America through a longitudinal study of 19 countries (1985-1995). A multiple regression analysis identifies income, political democracy, changes in the regulatory environment (including privatization) and the growth of cellular phones as the most relevant factors accounting for the expansion of fixed telephone networks. These findings agree with previous assessments (Mody & Tsui, 1995) that support the idea that competition and the presence of an independent regulator, rather than a simple shift from public to private ownership, are key to network expansion.

For countries with populations historically deprived of telecommunication services, the idea of overall network expansion seems to fulfill the promises brought about by liberalization. However, after more than a decade of liberal policies, almost two thirds of Latin American households do not have fixed telephones yet, and the penetration of mobile and fixed phones combined only covered 50% of the population (ITU, 2000). The region has become a major puzzle. It has hardly any unmet demand of telecommunication services[3], and yet, there are low rates of service penetration. A considerable 'depressed' demand seems to be one of the unwanted consequences of liberalization in the region (ITU, 2000:3). Rising local access prices, fostered in the context of tariff rebalancing, threaten to exclude a good part of the population.

---

[3] Conventional statistics of telephone demand, as revealed in waiting lists reported by operators, estimate 92% of demand in Latin America has been satisfied. (ITU, 2000:89)

As Milne notes (2000), tariff rebalancing is a necessary condition for an industry undergoing severe changes in its market structure. On the one hand, in competitive markets prices must reflect underlying costs, and basic services should operate without cross-subsidies received in the past from long distance or other services. On the other hand, technological innovations have affected the cost structure of the business, and incumbent companies need to adjust their tariffs to patterns that made them competitive. Even in those countries with closed markets for basic services, monopolies face competitive pressures of call back services, mobile cellular phones and IP telephony that become substitutes of fixed telephony.

Price rebalancing has tended to raise residential rentals facing operators and regulators with the dilemma of balancing sustainability and affordability of the service[4]. The tensions between the two conflicting goals are higher in low and middle-income countries where scarce economic sources limit the ability of maneuver of market agents. Testing a model that relates telephone service affordability to income distribution, Milne found that in low and middle-income nations price restructuring leading to significant increase in residential charges, "risks driving some subscribers off the network, and can slow down the rate of network expansion" (2000:919). It is important to assess how issues of pricing relate to service penetration. Do they support or hinder universal service goals? What alternative solutions are giving regulator of less developed countries to the tensions between liberalization and service affordability?

By comparing trends of telephone cost and telephone service penetration in Argentina, Brazil, Chile, Mexico, Peru and Venezuela during the last decade, this paper discusses how different approaches to telecommunication reform have fostered higher levels of access in developing countries during the transition from state monopolies to total liberalization. The discussion is framed by a survey of the universal service policies put in place by the six countries, and an analysis of their performance indicators. The goal is to draw lessons that would help to improve the regulatory strategies of telecommunication administrators, and to discuss the implications for new forms of governance defined in more liberalized scenarios. The key question is, have these changes contributed, and if so, in what ways, to more balanced and fairer systems of communication in developing countries?

---

[4] Reviewing preview studies on effects of tariff rebalancing, Milne (2000) concludes that rebalancing inevitably favors some customer groups over others, in particular heavy users (typically business and high income groups). However, outcomes of the evaluations differ from country to country, or between social groups. The author hypothesizes that income levels of the nation or community make a difference.

## METHODOLOGY

♦ **Design of the study**

      The analysis focuses on the elaboration of country profiles that identify the nation's universal service policy as defined in contracts and bylaws designed by telecommunication administrators. This study follows Hudson's (2000) proposed model for evaluation of universal access in developing nations presenting the evolution of different performance indicators over an eleven-year period. Finally, a comparative assessment discusses the ways in which universal service regulations have unfolded in the studied countries relating them to new forms of governance brought by liberal reforms.

♦ **Data**

      In 1997, the International Telecommunication Union started to keep track and report trends in telecommunication reform worldwide through the Sector Reform Unit (SRU), an arm of the Telecommunication Development Bureau (BDT). The work generated by the SRU included country profiles available on the World Wide Web. These data as well as the latest assessment of the Inter-American Telecommunication Commission (CITEL, 2000) on universal service are used as primary data sources. An evaluation of service penetration by country is supported with indicators gathered from the ITU's Stars database (2000) and ITU's Americas Telecommunication Indicators (2000).

♦ **Operational definitions**

      **Universal Service and Universal Access**. In line with the redefinition of universal service to include more than the wired telephone, this study looks at advances in <u>universal service</u> goals through two different indicators: (1) the percentage of households with a telephone line[5]; and (2) telephone penetration defined as the sum of dedicated lines plus wireless lines (mobile cellular) per 100 inhabitants. Both indexes combined are considered as a unique indicator of telephone penetration.

      Progress in <u>universal access</u> is measured as the number of payphones per 1000 inhabitants. Although the indicator does not account for rural versus urban penetration, it serves as a reference of the number of people that may have access to a neighboring telephone. Public telephony as a percentage of the overall fixed-line network can be used as a parameter to measure commitment that administrators and operators may have to spread the service.

      **Liberalization.** This concept is operationalized as the year in which the supply of basic telephone service is liberalized through the introduction of competition in local or long distance services, or through the introduction of competition in products which substitute for the local service, such as mobile cellular services.

---

[5] This indicator is estimated using a proxy based on number of residential lines as a percentage of households. Data was gathered from the ITU Start database.

**Service affordability**. The major components of pricing of the fixed telephony are installation charge, a one-time charge for new user, and the monthly subscription charge, applicable to all users. This study includes data on price of line installation as percentage of annual average per capita income, and monthly connection charge as percentage of monthly average per capita income. Same indicators are presented for mobile cellular telephony as a way to observe the competitive pressures played out in the process of tariff rebalancing. Data for this exercise is gathered from the ITU Start database. Some limitations for comparison arise from the lack of available information for all the eleven points in times.

## COUNTRIES OVERVIEW

### Argentina

In 1990 the Empresa Nacional de Telecomunicaciones (ENTEL), a state owned monopoly, was privatized and split into two separate regional companies that enjoyed a seven-year monopoly on basic services with a possible extension for 10 years based on satisfying contract requirements [i]. These requirements were laid out in 1991 in terms of the amount of investment to be made in the whole network, without specifying areas or services to be included in USO. The target was to achieve an average annual growth rate of 6% for the next 5 years. Government oversight was limited to determining the quality of operation. A newly created regulatory agency was mandated to guarantee, among other things, universal service. Operators assumed the expressed commitment to serve 400 areas in the northern region, and 280 in the southern region through public and semi-public service plans, and licenses were granted to independent operators who had been handling areas not covered by the public enterprise (CITEL, 2000). In spite of the mostly competitive environment, lack of interconnection regulations retarded the full operation of these licensed carriers during these first years (Straubhaar et al., 1995).

From the outset of the reform, competition was allowed in value-added services, private networks, satellite transmissions and mobile cellular services. Cellular operators were organized into regional duopoly markets. The main two competing companies were Movicom and Movistar. The first began operations in the country in 1989 when a consortium led by BellSouth was awarded a license to introduce the service in the Buenos Aires metropolitan area. In line with the 1990 privatization plan, Telefónica and Telecom were subsidized to compete in this market. However, in 1993 both companies became partners in the creation of Movistar and its service: the Miniphone. This association lasted until October 1999 when the joint venture was closed due to differences between the partners. Subscribers were then distributed between two companies: TCP and Telecom Personal. Table 2 presents the evolution of telephone penetration in Argentina. In this liberalized environment, fixed lines kept growing over the last decade. This growth has slowed down while mobile telephony reached record rates of expansion.

**Table 2. Argentina- Telephone penetration**

| ARGENTINA | | | Reform | | | | | | | | | | |
|---|---|---|---|---|---|---|---|---|---|---|---|---|---|
| | 1988 | 1989 | 1990 | 1991 | 1992 | 1993 | 1994 | 1995 | 1996 | 1997 | 1998 | 1999 | 2000 |
| Phonelines per 100 inh. | 10.0 | 9.6 | **9.3** | 9.5 | 10.8 | 11.8 | 13.7 | 15.9 | 17.4 | 18.8 | 19.7 | 20.1 | 21.5 |
| Cell.Pho. per 100 inh. | 0.0 | 0.0 | **0.0** | 0.1 | 0.1 | 0.3 | 0.7 | 1.0 | 1.6 | 4.5 | 7.0 | 12.1 | 12.2 |
| **Telephone Penetration** | 10.0 | 9.6 | **9.3** | 9.6 | 10.9 | 12.1 | 14.4 | 16.9 | 19.0 | 23.2 | 26.7 | 32.2 | 33.7 |

Source: ITU Stars Database

One of the major achievements in Argentina has been a consistent increase of accessibility and availability of telephone services (Table 3). Household penetration grew at average annual rates of 10% with a marked deceleration of this tendency after the target year of 1996, when the process became markedly more competitive. Expansion of public telephony indicates a compromise to serve social needs.

**Table 3. Argentina - Accessibility and availability**

| ARGENTINA | | | Reform | | | | | | | | |
|---|---|---|---|---|---|---|---|---|---|---|---|
| | 1988 | 1989 | 1990 | 1991 | 1992 | 1993 | 1994 | 1995 | 1996 | 1997 | 1998 |
| % of residential lines | 74.9 | 75.6 | **76.0** | 78.0 | 80.0 | 81.0 | 81.0 | 83.0 | 82.3 | 85.0 | 83.8 |
| Residential phones/ 100 households | 28.7 | 27.8 | **27.1** | 28.5 | 32.7 | 36.2 | 41.8 | 49.4 | 53.6 | 59.3 | 61.0 |
| | | | | | | | | | | | |
| % Payphones/main lines | 0.8 | 1.0 | **0.8** | 0.8 | 1.0 | 1.1 | 1.2 | 1.3 | 1.3 | 1.4 | 1.5 |
| Payphones per 1000 inh. | 0.8 | 0.9 | **0.7** | 0.8 | 1.1 | 1.3 | 1.7 | 2.0 | 2.3 | 2.6 | 3.0 |

Source: ITU Stars Database

However, the major change experienced at the end of the first phase of liberalization was the tremendous growth of mobile telephony. The major factor accounting for this growth was the introduction of the Calling Party Pays (CPP), a system whereby the caller is charged for making the call. This tariff scheme, along with the steady decrease of relative cost in subscription costs (Table 4), boosted mobile telephony increasing the penetration of the service 180% during two consecutive years (1997-1998).

**Table 4. Argentina- Telephone service affordability**

| ARGENTINA | | | | | | | | | |
|---|---|---|---|---|---|---|---|---|---|
| | 1990 | 1991 | 1992 | 1993 | 1994 | 1995 | 1996 | 1997 | 1998 |
| Fixed phone connection/ income per capita | 49.8 | 16.4 | 11.0 | 10.7 | 6.7 | 6.7 | 3.2 | 3.0 | 1.8 |
| Mobile phone connection/ income per capita | | 8.0 | 5.6 | 3.2 | 2.9 | 3.0 | | | |
| | | | | | | | | | |
| Res. mon. phone subc./ monthly per capita income | 4.0 | 1.3 | 1.4 | 1.4 | 1.3 | 1.4 | 1.4 | 1.9 | 1.9 |
| Res. mon. mobile subc./ monthly per capita income | | 7.2 | 5.8 | 5.7 | 5.3 | 5.5 | 4.8 | 4.5 | |

Source: ITU Stars Database

The restructuring stage, which guaranteed exclusivity to regional monopolies, was extended until 1999. In this period conceptualized as a transition toward total competition, licenses for basic service operators contained a detailed list of USO, which included: installation of semi-public long distance service in every community of 80 to 500 people; installation of wirelines in communities where at least 30 customers request the service; and new targets to enhance the public telephone network. In addition to these requirements in infrastructure development, in July 1999, the government passed General Regulations on Universal Services, which extended infrastructure commitments, and created three universal service categories: high-cost areas, specific customers and specific services. The actual plans recently designed to match these concepts aim at providing service to retirees, pensioners, low consumption customers, users with physical limitations, and educational and health institutions. The Regulations established that providers of telecommunication services must contribute to the Universal Service Fund with a percentage of total income earned from operations. The new legislation adopted a dynamic concept of universal service with periodic review.

**Brazil**

By the beginning of the 1990s, Telebrás was a holding company with 28 subsidiaries representing 27 local operators and a long distance carrier, Embratel. The holding served 91% of Brazilian fixed telephone subscribers. The remaining 9% was in the hands of four independent operators. Uneven development was the main characteristic of the Brazilian corporation by the late 1980s. Seventy-three percent of Telebrás' revenue came from Embratel and Telesp, the local operator of wealthy Sao Paulo state. Although the transition to a democratic regime reaffirmed the public monopoly in telecommunications in the new constitution of 1988, the new political scenario brought about changes to the state run corporation. Telebrás started to trade its shares in the stock market in 1989, creating a system in which the state sold its stock in the company, while retaining managerial control. By the mid-1990s, it was estimated that foreign investor held 36.7% of the equity (Worhlers de Almeida, 1998).

The Brazilian approach to reform was not privatization but the maintenance of state control over a corporation funded with public capital that faced competition in different markets. In August 1995 the Cardoso government launched the reform of public institutions including the telecom reform that began with the removal of the constitutional monopoly reserved to Telebrás. Priority was given to promote competition in the cellular telephone market, satellite telecommunication, data transmission, and other added-value services. Other important steps of the reform included tariff rebalancing, the merger of operators into regional groups, and the implementation of autonomous management structures. In 1997 the government awarded 10 licenses for mobile telephony, thus introducing competition into this market. Table 5 clearly shows that the State-run holding attained considerable improvements in telephone penetration before privatization.

**Table 5. Brazil- Telephone penetration**

| BRAZIL | 1989 | 1990 | 1991 | 1992 | 1993 | 1994 | Reform 1995 | 1996 | 1997 | 1998 | 1999 | 2000 |
|---|---|---|---|---|---|---|---|---|---|---|---|---|
| Phonelines per 100 inh. | 6.2 | 6.5 | 6.8 | 7.3 | 7.4 | 8.0 | **8.5** | 9.6 | 10.6 | 12.1 | 14.9 | 14.9 |
| Cell.Pho. per 100 inh. | 0.0 | 0.0 | 0.0 | 0.0 | 0.1 | 0.4 | **0.8** | 1.6 | 2.8 | 4.7 | 8.9 | 13.6 |
| **Telephone Penetration** | 6.2 | 6.5 | 6.9 | 7.3 | 7.6 | 8.4 | **9.3** | 11.1 | 13.4 | 16.7 | 23.8 | 28.5 |

Source: ITU Stars Database

In 1998 Telebras was divided and sold in a public bid establishing a duopoly system, which will operate until 2002. The government used privatization to attract foreign capital to finance the external deficit of the balance of payments, and to obtain fresh sources for the Treasury. Telebrás was broken up vertically and regionally into -12 sub holding companies and sold separately. Three regional operators offered a mix of wired telephony for local services and internal region long distance (the *Baby Braz*: T1/Telco North-Northeast/ T2, T2/Telco Center-South, and T3/Telesp); one operator for national and international long distance (Embratel); an eight spin-off creators supplying mobile telephony. The most profitable is Telesp, which serves Sao Paulo (ITU, 2000). The same year as privatization a new Decree on the General Plan of Universal Goals set USO binding license holders for the fixed telephone service. The plan sets goals of individual and collective access, and pays special attention to educational and health institutions with demands for advanced services and to individuals with special needs. Under this environment, telephone accessibility and availability continued to advance at a good pace (Table 6).

**Table 6. Brazil- Accessibility and availability**

| BRAZIL | 1988 | 1989 | 1990 | 1991 | 1992 | 1993 | 1994 | Reform 1995 | 1996 | 1997 | 1998 |
|---|---|---|---|---|---|---|---|---|---|---|---|
| % of residential lines | 70.0 | 69.0 | 69.5 | 70.0 | 70.0 | 69.0 | 67.9 | **67.6** | 68.1 | 69.0 | n/a |
| Residential phones/ 100 households | 16.8 | 17.0 | 17.8 | 18.7 | 19.8 | 20.0 | 21.1 | **22.4** | 25.3 | 28.6 | 28.6 |
| | | | | | | | | | | | |
| % Payphones/main lines | 2.6 | 2.5 | 2.4 | 2.3 | 2.4 | 2.6 | 2.6 | **2.6** | 2.7 | 2.8 | 2.8 |
| Payphones per 1000 inh. | 1.5 | 1.6 | 1.6 | 1.6 | 1.7 | 2.0 | 2.1 | **2.2** | 2.6 | 3.0 | 3.0 |

Source: ITU Stars Database

One of the major achievements of the Brazilian strategy was the tariff rebalancing process. The Cardoso government radically adjusted the structure of the price system. The result was a cut in installation charges from a peak of 1,500 US dollars in 1992 to 50 US dollar in 1998 (ITU, 2000). The adjustment combined with the adjustment of the Brazilian currency and a better overall economic performance drove the relative costs of the service down (Table 7) .

**Table 7. Brazil- Telephone service affordability**

| BRAZIL | 1994 | 1995 | 1996 | 1997 | 1998 |
|---|---|---|---|---|---|
| Fixed phone connection/ income per capita | 49.2 | 26.9 | 22.7 | 1.5 | 0.9 |
| Mobile phone connection/ income per capita | 8.6 | 8.0 | 6.7 | 6.1 | |
| | | | | | |
| Res. mon. phone subc./ monthly per capita income | 0.2 | 0.8 | 0.7 | 1.7 | 1.7 |
| Res. mon. mobile subc./ monthly per capita income | 12.5 | 7.9 | 6.6 | 6.0 | |

Source: ITU Stars Database

**Chile**

For most of the 20[th] century the Chilean telecommunication system evolved as a private enterprise[ii]. During the period of state control (1973-1986), the national system was reorganized into two regional monopolies for local services, CTC and ENTEL. Competition in long distance was introduced in 1986, the same year that key components of CTC and ENTEL were privatized. During the first period of liberalization (1986-1993) the government set no USO. This left the pace of network expansion to market forces. In this environment, telephone density grew at a healthy annual average rate of 16%. In spite of the strong growth of mobile and fixed telephony (Table 8), by the early 1990s the network was showing two major flaws: a virtual halt of service growth in rural areas, and the decrease of payphone penetration (SUBTEL, 2000). Figures in Table 10 clearly depict the problem. The situation also revealed the flaw in having a weak regulatory agency with few legal tools at its disposal to deal with the problem.

**Table 8. Chile- Telephone penetration**

| CHILE | | | | Reform | | | | | | | | | | | |
|---|---|---|---|---|---|---|---|---|---|---|---|---|---|---|---|
| | 1985 | 1986 | 1987 | 1988 | 1989 | 1990 | 1991 | 1992 | 1993 | 1994 | 1995 | 1996 | 1997 | 1998 | 1999 | 2000 |
| Phone lines per 100 inh. | 4.4 | 4.5 | 4.6 | 4.9 | 5.0 | 6.6 | 7.9 | 9.5 | 11.0 | 11.3 | 12.7 | 14.9 | 16.1 | 18.6 | 20.7 | 22.1 |
| Cell. Pho. per 100 inh. | 0.0 | 0.0 | 0.0 | 0.0 | 0.0 | 0.1 | 0.3 | 0.5 | 0.6 | 0.7 | 1.4 | 2.2 | 2.8 | 6.5 | 15.1 | 22.4 |
| Telephone Penetration | 4.4 | 4.5 | 4.6 | 4.9 | 5.0 | 6.7 | 8.2 | 9.9 | 11.7 | 12.1 | 14.1 | 17.1 | 18.9 | 25.1 | 35.8 | 44.5 |

Source: ITU Stars Database

**Table 9. Chile - Accessibility and availability**

| CHILE | | | | Reform | | | | | | | | | |
|---|---|---|---|---|---|---|---|---|---|---|---|---|---|
| | 1985 | 1986 | 1987 | 1988 | 1989 | 1990 | 1991 | 1992 | 1993 | 1994 | 1995 | 1996 | 1997 | 1998 |
| % of residential lines | 68.0 | 70.4 | 70.4 | 72.0 | 72.3 | 75.5 | 75.5 | 76.9 | 77.0 | 77.0 | 77.2 | 77.0 | 77.0 | 77.0 |
| Residential phones/ 100 households | 13.8 | 14.5 | 14.9 | 16.1 | 16.4 | 22.4 | 27.0 | 33.4 | 39.0 | 40.1 | 45.3 | 52.6 | 56.6 | 66.2 |
| | | | | | | | | | | | | | | |
| % Payphones/main lines | 1.8 | 1.8 | 1.9 | 1.9 | 2.1 | 2.0 | 1.6 | 1.5 | 1.2 | 0.6 | 0.6 | 0.5 | 0.5 | 0.5 |
| Payphones per 1000 inh. | 0.8 | 0.8 | 0.9 | 0.9 | 1.0 | 1.3 | 1.3 | 1.4 | 1.3 | 0.7 | 0.8 | 0.8 | 0.8 | 0.9 |

Source: ITU Stars Database

In 1994, an amendment to the law established a Development Fund with contributions from long distance service providers. This was aimed at subsidizing the expansion of public telephones in marginal, low-income and rural areas. The reform privileged universal access goals over universal service provision. The standard for universal access formulated by law mandated at least one payphone in any community with more than 60 inhabitants, located at least half an hour's driving from the next nearest telephone. Preliminary estimates indicate that at the current rate of expansion, 80% of the rural population will be covered by the year 2002 (SUBTEL, 2000). In 1999 an important amendment to the law changed the original goal of "payphone" to "telecommunication services". Although the ramifications of the shift are as yet unclear, it may suggest a move away from universal access to universal service definition. The Chilean Secretariat of Telecommunications has acknowledged the need for adding rules that boost access to telecommunication services, mainly in the case of lower income groups of the population.

Over the 1990s, the affordability of the telephone service in Chile was guaranteed by the sustained improvement of the economy as a whole. Chile has one of the world's cheapest tariffs systems for long distance service but the installation costs and rates for local calls (11 US cents per minute) are above the regional average (ITU 2000). The consistent increase of the overall income level keep the relative cost of the service down making it affordable to the population (Table 10).

**Table 10. Chile- Telephone service affordability**

| CHILE | | | | | | |
|---|---|---|---|---|---|---|
| | 1993 | 1994 | 1995 | 1996 | 1997 | 1998 |
| Fixed phone connection/ income per capita | 8.0 | n/a | 4.0 | 3.7 | 3.4 | 3.2 |
| Res. mon. phone subc./ monthly per capita income | 5.68 | n/a | 5.15 | 4.52 | 4.17 | 3.98 |

**Mexico**

In 1990 a consortium led by Grupo Carso from Mexico, in partnership with BellSouth and France Telecom, bought 20% of Teléfonos de México (TELMEX) stock, thus acquiring rights to operate a monopoly of basic services until 1996 [iii]. Once laws removing caps on foreign investment were lifted, the government then also sold its holding to international investors. Regional duopolies were established for mobile cellular services. TELMEX was able to participate in all these markets through its subsidiary -TELCEL- (Straubhaar at al., 1995). During this period, TELMEX was required: a) to achieve an average annual growth of 12% in main lines through 1994, b) by 1994, to install payphones in each town with a population greater than 500, and c) to raise the payphone penetration up to 2 per 1,000 inhabitants in 1994, and 5 per 1,000 in 1998. The Telecommunication Act of 1995 laid out social coverage as an objective for public networks. The mandate was formulated through new plans for rural and public telephony, providing cellular facilities to grocery stores, health centers and mobile satellite connection to medical services (ITU 2000).

In 1996, the long distance market was opened to competition while the basic service continued to be developed under regional duopolies through the entrance of other providers that competed with TELMEX. Expansion of fixed lines has kept conservative rates while mobile telephony has double after the end of the exclusivity period.

**Table 11. Mexico- Telephone penetration**

| MEXICO | | | | Reform | | | | | | | | | |
|---|---|---|---|---|---|---|---|---|---|---|---|---|---|
| | 1987 | 1988 | 1989 | 1990 | 1991 | 1992 | 1993 | 1994 | 1995 | 1996 | 1997 | 1998 | 1999 | 2000 |
| Phonelines per 100 inh. | 5.1 | 5.3 | 5.8 | **6.5** | 6.8 | 7.5 | 8.4 | 9.2 | 9.4 | 9.3 | 9.7 | 10.4 | 11.2 | 12.5 |
| Cell.Pho. per 100 inh. | 0.0 | 0.0 | 0.0 | **0.1** | 0.2 | 0.3 | 0.4 | 0.6 | 0.7 | 1.1 | 1.8 | 3.5 | 7.8 | 14.2 |
| **Telephone Penetration** | 5.1 | 5.3 | 5.8 | **6.6** | 7.0 | 7.9 | 8.8 | 9.8 | 10.1 | 10.4 | 11.5 | 13.8 | 19.0 | 26.7 |

Source: ITU Stars Database

By that year the company had only partially fulfilled USO established by the concession. The annual growth target was attained during the first four years of privatization, but network expansion dropped to 4% in 1994 and 5% in 1995. In 1996, it experienced zero-growth. At the same time, the increase in overall telephone penetration was not comparable to the growth in household penetration suggesting that the expansion was driven by commercial demand. TELMEX met the 1994 target in payphones. But the number of public phones declined in 1996. And TELMEX fell far short of fulfilling its 1998 requirement (see Table 12).

**Table 12. Mexico - Accessibility and availability**

| MEXICO | | | | Reform | | | | | | | |
|---|---|---|---|---|---|---|---|---|---|---|---|
| | 1987 | 1988 | 1989 | 1990 | 1991 | 1992 | 1993 | 1994 | 1995 | 1996 | 1997 | 1998 |
| % of residential lines | 69.0 | 70.0 | 69.9 | **71.0** | 70.9 | 71.7 | 71.0 | 71.0 | 74.1 | 74.6 | 74.6 | n/a |
| Residential phones/ 100 households | 19.3 | 20.3 | 21.8 | **23.7** | 25.6 | 28.2 | 30.6 | 33.2 | 35.3 | 33.9 | 34.5 | 34.5 |
| | | | | | | | | | | | | |
| % Payphones/main lines | 1.1 | 1.2 | 1.3 | **1.6** | 1.6 | 1.9 | 2.3 | 2.6 | 2.8 | 2.7 | 3.2 | 3.2 |
| Payphones per 1000 inh. | 0.6 | 0.6 | 0.8 | **1.0** | 1.1 | 1.4 | 2.0 | 2.3 | 2.6 | 2.5 | 3.1 | 3.3 |

Source: ITU Stars Database

An important flaw in the Mexican strategy stemmed from the lack of population data needed to estimate the coverage of the payphone network. By 1998, TELMEX had provided the service to 16,000 localities with more than 500 inhabitants. However, most rural towns have less than 500 inhabitants leaving almost 10 million people (10% of the population) out of expansion plans for public telephony (ITU, 1998b).

Facing economic pressures during the peso crisis, TELMEX held expansion plans back. However, over the same period, mobile cellular service penetration was growing by rates between 20% and 90% per year. The growth of mobile telephony was based on two pricing strategies: CPP and pre-paid plans. Although the tariff rebalancing process situated fixed phone in a competitive position (Table 13), by the end of 1999 85% of TELMEX cellular subscribers were using the pre-paid plan "Amigo" (ITU, 2000). Cofetel, the Mexican regulator, currently reviews the CPP system to determine if revenue split (20/5 US cents) between mobile and fixed operators reflects the costs of both operators. The underlying question is if a system that sets advantages to mobile users is in fact discriminating against subscribers of fixed-lines. As the price of mobile services continues falling the disadvantage for fixed-phone users will become more evident.

**Table 13.- Mexico- Telephone service affordability**

| MEXICO | 1989 | 1990 | 1991 | 1992 | 1993 | 1994 | 1995 | 1996 | 1997 | 1998 |
|---|---|---|---|---|---|---|---|---|---|---|
| Fixed phone connection/ income per capita | 9.9 | 15.5 | 10.8 | 11.9 | 12.0 | 11.7 | 9.2 | 6.8 | 2.9 | 2.5 |
| Mobile phone connection/ income per capita | | 16.3 | 3.7 | 3.5 | 3.2 | 0.6 | 0.6 | 0.4 | | |
| | | | | | | | | | | |
| Res. mon. phone subc./ monthly per capita income | 1.1 | 1.8 | 2.7 | 2.7 | 3.0 | 3.2 | 2.9 | 3.3 | 3.9 | 3.9 |
| Res. mon. mobile subc./ monthly per capita income | | 13.4 | 13.9 | 13.2 | 12.0 | 12.1 | 10.7 | 10.0 | 8.6 | 7.2 |

Source: ITU Stars Database

Recent accounts of the post-liberalization period have shown that USO constituted a major point of contention between TELMEX and competitors in disputes over interconnection and regulatory transparency (Burkart, 2000; Gómez-Mont 2000). By 1999, the regulatory authority had not identified a universal service program that balanced the interests of all providers in the market. TELMEX wanted to have USO reduced, eased or shared with competitors. However it was also ready to use USO as an argument to justify high interconnection charges. As Burkart notes, "with so little put into expanding the [network], either through interconnection or through and aggressive universal service plan, the development of markets of basic services cannot proceed." (2000:235)

**Peru**

In May 1994 Telefónica of Spain bought 35% of Telefónica del Peru (TP), a newly created consortium that consolidated local and long distance operators, Entel and CPT, into one national company. The transaction ranked TP as the company with the highest cost per line in the region. The outcome surprised observers who considered that the four-year exclusivity period awarded to the company was particularly short in comparison with previous privatizations (e.g. Mexico, Venezuela). The high value of TP stemmed from market power guaranteed in the monopoly franchise -including mobile telephony-, which in fact promoted the vertical integration of the holding and reducing competitive pressures of product substitute (mobile telephony, see Table 11) during the first years of the privatization (Briceño, 1999).

**Table 14. Peru- Telephone penetration**

| PERU | Reform | | | | | | | | | | | | | | |
|---|---|---|---|---|---|---|---|---|---|---|---|---|---|---|
| | 1986 | 1987 | 1988 | 1989 | 1990 | 1991 | 1992 | 1993 | 1994 | 1995 | 1996 | 1997 | 1998 | 1999 | 2000 |
| Phone lines per 100 inh. | 2.2 | 2.2 | 2.4 | 2.5 | 2.6 | 2.5 | 2.7 | 3.0 | 3.3 | 4.7 | 6.0 | 6.8 | 6.3 | 6.7 | 6.4 |
| Cell. Pho. per 100 inh. | 0.0 | 0.0 | 0.0 | 0.0 | 0.0 | 0.0 | 0.1 | 0.2 | 0.2 | 0.3 | 0.8 | 1.8 | 3.0 | 4.0 | 4.0 |
| Telephone Penetration | 2.2 | 2.2 | 2.4 | 2.5 | 2.6 | 2.5 | 2.8 | 3.1 | 3.5 | 5.0 | 6.8 | 8.5 | 9.3 | 10.7 | 10.4 |

The monopoly franchise established goals of network expansion (250,000 new lines by the year 1999), and provisions for the installation of at least one telephone in 1,540 rural settlements with population over 500 inhabitants. Under the framework of the franchise commitments, TP agreed with the regulatory agency, Osiptel, on a program of telecenters developed in partnership with small entrepreneurs aimed at providing public telephony to low-income areas. In 1995 the plan was enhanced through the program of *Cabinas Públicas*, telecenters that offer Internet access. The Scientific Peruvian Network (Red Científica Peruana) opened the first of these centers in 1995. Since that year, the telecenter network has continuously grown becoming a prosperous network of 1,000 sites. The demand is driving prices down and the cost of one hour of computer/Internet connection currently varies between 0.7 and 0.85 US dollar (Proenza et al., 2000).

**Table 15. Peru - Accessibility and availability**

| PERU | | | | | Reform | | | | | |
|---|---|---|---|---|---|---|---|---|---|---|
| | 1989 | 1990 | 1991 | 1992 | 1993 | 1994 | 1995 | 1996 | 1997 | 1998 | 1999 |
| % of residential lines | 72.6 | 70.0 | 70.0 | 70.6 | n/a | n/a | n/a | 82.0 | 82.0 | n/a | n/a |
| Residential phones/ 100 households | 9.6 | 9.6 | 9.0 | 10.1 | n/a | n/a | n/a | 25.0 | 28.1 | 28.1 | n/a |
| | | | | | | | | | | | |
| % Payphones/main lines | 0.0 | 1.3 | 1.5 | 1.5 | 1.1 | 1.5 | 2.5 | 2.6 | 2.4 | 3.0 | 3.6 |
| Payphones per 1000 inh. | 0.0 | 0.3 | 0.4 | 0.4 | 0.3 | 0.5 | 1.2 | 1.5 | 1.6 | 1.9 | 2.4 |

Source: ITU Stars Database

At the end of the monopoly (1999), the major success of the reform in Peru was the strong growth of public access (see Table 15). Household penetration and fixed phone expansion has stagnated since 1996. A somehow repressed demand seemed to have jumped on the bandwagon of mobile telephone expansion. As with the case of Argentina, the introduction of CPP systems in 1996, combined with pre-paid plans, triggered demand for cellular phones, which already enjoyed a more competitive price structure (Table 16). The reform fulfilled the goal of tariff rebalancing of fixed telephony, but during the period of adjustment, the demand moved to the cheaper option available in the market. By the end of 1998 TP introduced the "Popular Telephone", a pre-paid plan for fixed subscribers that a year later accounted for 14% of all TP's customers (ITU, 2000). The experience speaks of the need and opportunities for options that reach a demand restricted by economic difficulties.

**Table 16. Peru- Telephone service affordability**

| PERU | | | | | | | | | |
|---|---|---|---|---|---|---|---|---|---|
| | 1990 | 1991 | 1992 | 1993 | 1994 | 1995 | 1996 | 1997 | 1998 |
| Fixed phone connection/ income per capita | 27.1 | 17.1 | 17.3 | 32.2 | 25.2 | 20.0 | 17.1 | 13.1 | 5.9 |
| Mobile phone connection/ income per capita | | | | | | | | | |
| | | | | | | | | | |
| Res. mon. phone subc./ monthly per capita income | 4.5 | 1.0 | 0.9 | 2.0 | 3.6 | 4.3 | 5.7 | 7.3 | 7.0 |
| Res. mon. mobile subc./ monthly per capita income | | | | | 26.8 | 22.9 | 11.6 | 12.2 | |

**Venezuela**

The privatization of the C.A. Teléfonos de Venezuela (CANTV) was part of the reorganization of state enterprises, a plan inscribed into major economic reforms undertaken by Venezuela in 1989 [iv]. A consortium led by GTE won the auction of 40% of the shares by the end of 1991. Eleven percent of the stock was allocated to company workers. The remaining 49% of stock originally in the hands of the State, has been partially sold in national and international stock markets. Today 89% of CANTV's equity is privately held.

CANTV monopolized local and long distance services for 10 years, with the exception of public telephony, a business which the government allowed new entrants into to supply service in areas not covered by CANTV. Five years after privatization, the government had licensed 30 firms in public telephony and 3 for rural areas. But long discussions over interconnection fees and what constituted "not attended" areas delayed the growth of competition in these segments. Since 1990, the mobile cellular market has operated as a duopoly, in which CANTV competes through a subsidiary (TELCEL). Full competition prevails in private networks and value-added services.

The concession for the basic service monopoly committed CANTV to meet goals of quality and service expansion specified by regions. The requirements included the addition of 355,000 new lines per year between 1992 and 2000, with fixed caps for new public telephones. After 1995, making use of contractual provisions, CANTV renegotiated these requirements arguing that the recession in the Venezuelan economy since 1993 has drastically altered the scenarios for demand originally envisaged in the concession. Conatel, a sector regulator that has been shaken by constant changes in its administration, has led the renegotiation of terms of the concession. As in Mexico, economic recession impacted heavily on targets for telephone service penetration, an indicator that has experienced negative growth in recent years (Table 17).

**Table 17. Venezuela- Telephone service penetration**

| VENEZUELA | | | | Reform | | | | | | | | |
|---|---|---|---|---|---|---|---|---|---|---|---|---|
| | 1988 | 1989 | 1990 | 1991 | 1992 | 1993 | 1994 | 1995 | 1996 | 1997 | 1998 | 1999 | 2000 |
| Phonelines per 100 inh. | 7.9 | 7.8 | 8.2 | **8.1** | 9.0 | 10.0 | 10.9 | 11.4 | 11.7 | 12.2 | 11.7 | 10.9 | 10.9 |
| Cell.Pho. per 100 inh. | 0.0 | 0.0 | 0.0 | **0.1** | 0.4 | 0.9 | 1.5 | 1.9 | 2.6 | 4.7 | 8.7 | 14.3 | 14.4 |
| **Telephone Penetration** | 7.9 | 7.8 | 8.3 | **8.2** | 9.3 | 10.8 | 12.4 | 13.2 | 14.3 | 16.9 | 20.3 | 25.2 | 25.3 |

Source: ITU Stars Database

However, unlike the Mexican case, household service penetration in Venezuela has increased at a faster pace (Table 18). Waiting lists, which were mainly composed by residential customers, has disappeared. This trend, in the context of decreasing fixed phone penetration was accompanied by a very strong expansion of mobile cellular services, a phenomenon that indicates the migration of the demand for basic services from wired networks to wireless services.

**Table 18. Venezuela - Accessibility and availability**

| VENEZUELA | | | | Reform | | | | | | |
|---|---|---|---|---|---|---|---|---|---|---|
| | 1988 | 1989 | 1990 | 1991 | 1992 | 1993 | 1994 | 1995 | 1996 | 1997 | 1998 |
| % residential lines | 70.0 | 68.3 | 68.3 | **69.0** | 68.0 | 69.0 | 67.6 | 67.9 | 68.6 | 70.0 | 71.1 |
| Residential phones/ 100 households | 30.3 | 28.1 | 26.3 | **30.7** | 33.5 | 37.7 | 40.5 | 41.8 | 44.7 | 47.8 | 45.9 |
| | | | | | | | | | | | |
| % Payphones/main lines | 2.1 | 2.1 | 2.1 | **2.0** | 2.5 | 2.4 | 2.4 | 2.3 | 2.1 | 2.5 | 2.8 |
| Payphones per 1000 inh. | 1.6 | 1.7 | 1.7 | **1.6** | 2.2 | 2.4 | 2.6 | 2.6 | 2.5 | 3.0 | 3.2 |

Source: ITU Stars Database

Today, Venezuela is one of twelve telecom systems worldwide where there are more mobile subscribers than fixed lines in use. Various factors account for the phenomenon. First, the political tension and turmoil that characterized Venezuela during the past decade frustrated efforts by the regulator at different times to rebalance tariffs. Second, the economic performance of the country never kept up with the process of devaluation of the Venezuelan currency. As a result, fixed telephone charges have become more expensive in relative terms (Table 19). Third, aggressive competition in the market of mobile telephony was based on the implementation of pre-paid plans accounting for the 73% of the subscriber base (ITU, 2000).

**Table 19. Venezuela- Telephone service affordability**

| VENEZUELA | 1990 | 1991 | 1992 | 1993 | 1994 | 1995 | 1996 | 1997 | 1998 |
|---|---|---|---|---|---|---|---|---|---|
| Fixed phone connection/ income per capita | 1.2 | 1.5 | 1.3 | 1.5 | 1.4 | 0.8 | 2.0 | 1.9 | 2.4 |
| Mobile phone connection/ income per capita | | | | | | | | | 0.3 |
| Res. mon. phone subc./ monthly per capita income | 0.4 | 0.4 | 0.6 | 1.1 | 1.1 | 1.1 | 2.4 | 2.1 | 2.4 |
| Res. mon. mobile subc./ monthly per capita income | | | | 16.1 | 18.9 | 15.6 | | 9.2 | 9.9 |

Source: ITU Stars Database

In 2000 the government and Congress cooperated on setting up a new regulatory framework to open the market of basic services. A new Telecommunication Act was recently approved including the creation of a Universal Service Fund with contributions of for-profit telecommunication service operators of 1% of their annual gross income. The Act assumed an active concept of universal service, encompassing all available services in the market (including Internet access), and the government launched an e-government initiative, which required all regional and national authorities to offer electronic services by the year 2002.

## COMPARATIVE ANALYSIS OF TRENDS

Has liberalization delivered the promises set forth a decade ago? In some terms, it has exceeded them but in others it has fallen behind. Consistent with past studies (Mody & Tsui, 1995; Straubhaar et al., 1995;  Petrazzini & Clark, 1996) this review shows that liberalization, and more importantly competition, resulted in significant increases in availability and accessibility of telecommunication services diversifying access points and increasing options. Regardless of the approach embraced by telecommunication administrators, liberalization has served the goals of accelerating overall basic service penetration. However, countries are facing problems in maintaining or increasing network access at the same pace after the first phase of reform, typically about four years after competition is introduced. After the fifth year of liberalization, network growth levels off resulting in a plateau in household and payphone penetration. This trend is present regardless of the USO strategy embraced by the country. Curiously, countries with no USO (Chile) or with flexible USO (Argentina) exhibit a better performance of household penetration (Graph 1).

**Graph 1. Availability - Household telephone penetration (Phones/ 100 households)**

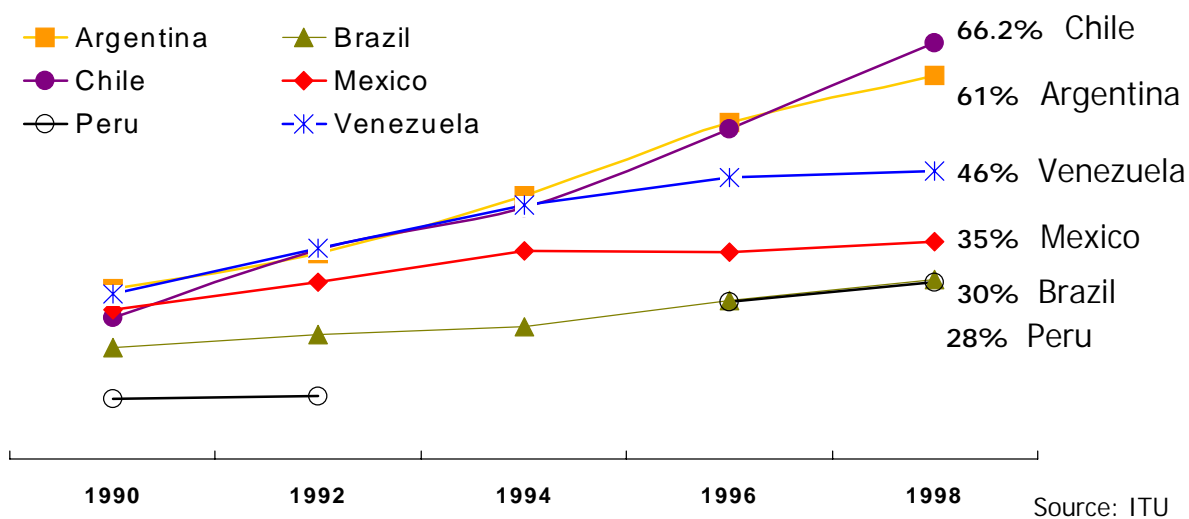

Source: ITU

In spite of the evidence, developing countries have bet more on privatization than on competition when coming to redefine their supply strategies. In their attempt to make the sale more attractive to buyers, less developed nations have awarded monopoly conditions to private operators while reducing regulatory oversight. Later, in the transition from monopoly conditions to market structures, administrators find themselves lacking the tools to align telecommunication policies with social and economic needs (Melody, 1999).

The main achievement of managed liberalization schemes, those that imposed USO on private monopolies (Brazil, Mexico, Peru, Venezuela), seems to be related to stronger rates of public telephone penetration (Graph 2). It may indicate a higher commitment for assuring service accessibility and meeting social goals. However, some of these cases also present a common problem: Mexico and Venezuela renegotiated and restated USO by lowering caps for expansion of the overall network, while focusing on plans of public telephony. One of the major obstacles to evaluating the real benefits of this strategy is the lack of data on rural versus urban penetration. Plans to expand private or public telephony as expressed in licenses, contracts and bylaws did not relate service targets to specific populations (low-income groups, disabled people). Operators do not report data on penetration in rural areas either to ITU databases or to CITEL records, limiting their plans for rural areas to public telephony.

**Graph 2. Accessibility - Payphones per 1,000 inhabitants**

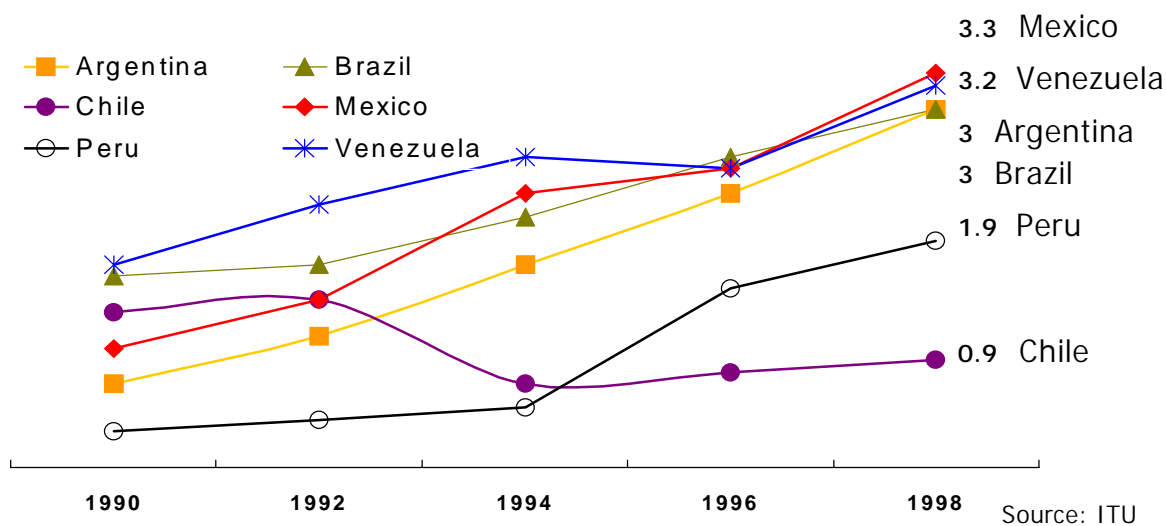

In trying to balance the contradictory interests of operators and the public, administrators have had to agree on more conservative scenarios for network expansion. The situation could certainly point out a tendency to wrongly estimate the parameters set forth in reform scenarios. It could also be evidence of the lack of an assertive administration in a more deregulated environment. In either case, it represents the need for flexible regulatory mechanisms that target specific groups in greater need.

Saturation of demand cannot be argued as a reason for the deceleration of growth. In countries with overall telephone penetration between 61% and 10.4%, universal service is still a pending goal. One reason for this could be a failure to deal effectively with issues of affordability and tariffs as the network progresses from one stage of development to another. Tariff rebalancing has reduced connection charges while driving monthly charge up (Graph 3). Venezuela, in particular, has been affected by increasing telephone charges, a factor that may explain the migration of users to mobile telephone services. In countries with severe problems of income distribution, and that have continuously faced economic difficulties, there is still a tremendous need for more socially desirable pricing schemes.

**Graph 3. Affordability -Tariff and income**

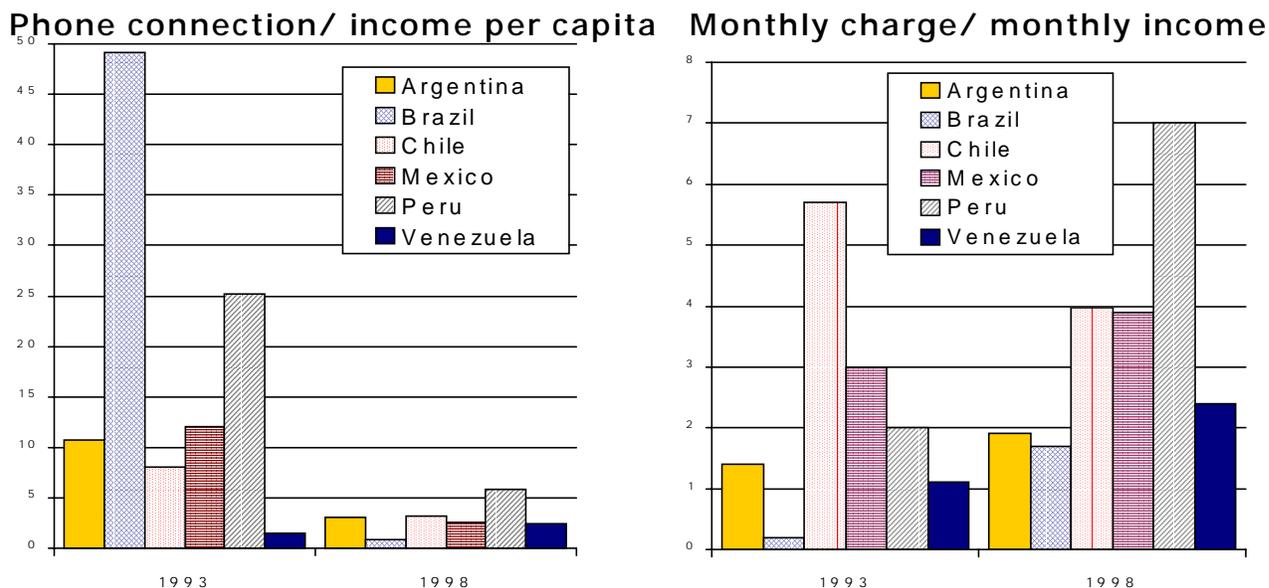

Source: ITU Start database

Experience with the massive use by mobile service providers of pre-paid plans as part of their marketing strategies shows the advantages and strong potential response to repressed demand. Fixed telephone providers in Peru (1998) and Venezuela (2000) have implemented similar plans discovering that low-income groups have more purchasing power than expected. The alternative emerged as a marketing strategy of the incumbents to cope with the competitive pressures of mobile telephone operators, but regulators remain inactive about this important issue.

The virtual stoppage in fixed-line network expansion should be directly related to the tremendous growth in mobile cellular services. In theory, mobile telephony should enhance the options for increasing service availability and accessibility. But in the case of some of these nations, cellular phones may not just become a supplement to a fixed phone but a substitute for it.

Although the ITU considers this scenario as a favorable one, from the industry point of view, the period of exclusivity represented higher economic pressures for the expansion of fixed-line networks. The majority of the reforms have set free cellular markets while keeping monopolies for basic services, as a way to assure to fixed phone operators revenues that guarantee profitability and network expansion. However, in countries with low telephone penetration and a high demand for service, monopolies moved slower than mobile cellular services in responding to demand.

Mobile cellular altered the equation by becoming a shortcut to meet people's needs. In countries with higher fixed-phone penetration, cellular operators seek subscribers more aggressively, offering lower prices, hence making the service more accessible to people. However, in countries with low telephone penetration the high demand is usually translated into higher prices for mobile services, setting them off as the preserve of higher income groups. In these nations the cellular market captures a highly profitable segment of the population. However, options of access are not enhanced for lower income groups.

The relevant indicator to observe the phenomenon is not teledensity, but substitution rates –the ratio of mobile cellular subscribers to total telephone subscribers, which indicates "the degree to which mobile cellular is used as an alternative rather than a supplement to fixed-line networks" (ITU, 1998b, p.49). A high ratio suggests that many cellular subscribers have no alternative for telephone service. Substitution rates are higher in Mexico, Brazil and Venezuela, that is countries with lower per capita income, whose fixed telephony had lower penetration rates at the moment of introduction of mobile services (see Graph 4).

**Graph 4.- Fixed/ Cellular Substitution**

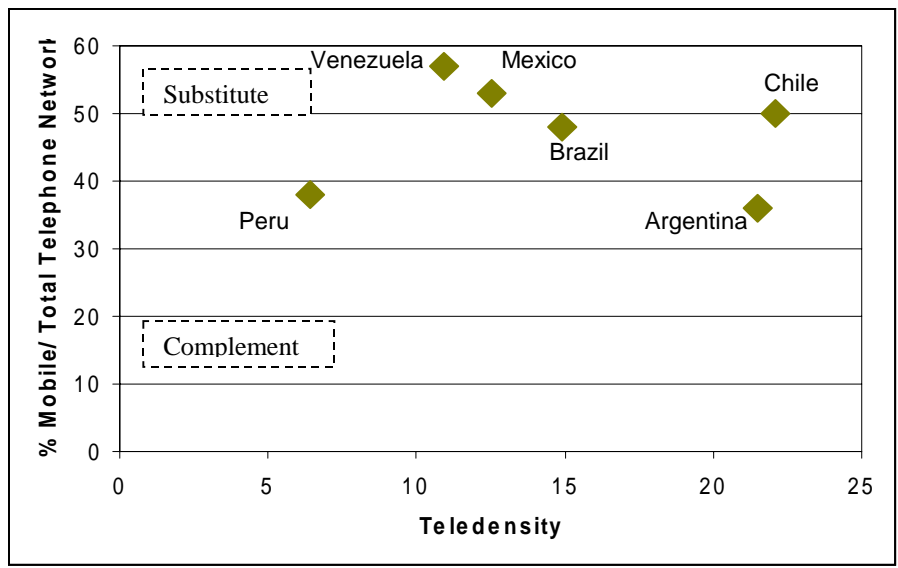

Venezuela is a good example of the trend being examined here. Since 1998 the number of subscribers to cellular phone services has exceeded subscribers of fixed telephones. Teledensity statistics register a negative growth of –4% , but the combined indicator of *telephone penetration* increased 21%. What type of challenge then does this phenomenon pose to regulators?

The need for a redistribution of resources in freer markets has led to a second round of legislative reforms. The first generation of USO, highly focused on general target of service accessibility, is being replaced by a second generation of USO. Chile (1994), Peru (1995), Mexico (1997), and more recently Argentina (1999), Brazil (2000) and Venezuela (2000) have started to explore this path through new legislation and the creation of Universal Service Funds. The establishment of public funds with contributions from competing telecommunication providers is a starting point to solve the Gordian knot of financing the incorporation of citizens deprived from the service. However, it is important to move beyond the supply-side discussion to consider the demand-side of who is effectively receiving the service and what means are more appropriate to enforce USO. The Mexican and Venezuelan cases suggest that more attention should be paid to alternatives that increase individual access, in particular taking advantage of the rapid growth of mobile cellular networks. Community solutions, such as telecenters, have also not been completely developed in governmental policy. With the exception of Peru and Chile, no other analyzed countries implemented public policies to bring communities into the effort of enhancing access. Venezuela and Brazil have recently launched national policies on public Internet access through telecenters.

## POLICY IMPLICATIONS

Mosco remind us that telecommunication reforms redefine the role of the state as settler of social claims. Drawing on political and social theory, he identifies four modes of state intervention relevant to telecommunication. They are representation, expertise, social control and market control. 'Representation' marks the degree of incorporation of the social in the decision making process from the point of view of a more deliberative formulation of regulations. 'Expertise' suggests an extremely narrowed procedure in regulation formulation carried out by boards of experts. 'Social control' and 'market control' are set off by the degree of direct or indirect control exerted upon the operations not only by the state but also by labor unions, other business sectors, and the public. Social control relies on corporatism while market control sees competition as the major form of governance. Based on this model, this study tries to identify shifts of modes of governance brought about by telecommunication reforms in five Latin American countries.

The shift from state corporatism to market competition models in the six studied countries has not gone beyond the '*expertise'* form of governance outlined by Mosco (1988), in which the power of newly created regulatory agencies remains captive to technical debates that represent the interest of a narrow group of providers with little concern for social claims. In this context, the 'hybrid' regulatory system of the transitions has hampered USO in countries that try to apply strict rules to monopolies of basic services, and a set of more flexible commitments to substitute services, such as mobile cellular, that operate in competition. The virtual failure of USO policies in Venezuela and Mexico should be interpreted in light of these contradictions.

Mansell and Wehn's (1998) international assessment of the progress of the information society suggests that in the deployment of 'the network of networks' that substitutes for the homogeneous network of PTTs, developing countries should combine three major strategies: (1) the increase of competitive pressures, (2) the use of all technologies available in the market for enhancing means of interconnection, and (3) the enforcement of USO. These authors argue that in this process led by technological convergence, regulation is "necessary to provide a foundation upon which markets can function more effectively than they could otherwise" (p.16). The *multi-network* perspective should be enhanced through a *multi-leveled* approach, a set of moving targets addressing the needs of access at the public, institutional and individual level (Hudson, 2000). All these elements demand a better and more active regulator, one capable of articulating innovative USO targeting users with special needs.

An alternative to moving away from the *expertise* model towards the path of *market control* is more and more effective competition in a diversified market. Multiplicity of providers increases the negotiation power of the regulator as settler of social claims. A good example of the step that should be taken in this direction is the strategy displayed by the Chilean and Brazilian authorities in dividing up the market and introducing competition along with privatization.

Once the network increases its points of access through different services, policy makers should take advantage of the reconfiguration fostered by competitive markets to meet social goals. Regulators need to develop a single definition of universal service that extends social commitments to all operators of the telecommunication market. This definition should cut across services without being bundled into technological specificity. It ought to incorporate factors from the consumer-demand perspective accounting for the urban-rural continuum, and social and economic strata. Peru, Chile, Argentina and Venezuela have adopted steps in this direction. But it is not yet clear how this new conceptualization of *basic telecommunication service* will get enacted through concrete plans.

Technological convergence is reshaping the market. Regulation is key to set in motion the process that uses the sum total of capabilities available in the market to meet the needs of specific segments of the population still deprived from telecommunication services. As Melody (1999) has argued, universal service regulations can provide an answer that bridges divergent interests and captures network externality benefits that competitive markets cannot achieve. This is possible through the incorporation of "unphoned" users who still may have resources to pay for the basic service, thus enhancing network externalities. However, this would demand a market-regulated system that attends to a wider range of social claims than is currently the case, by setting fair rules for all participants in the market, including consumers and newcomers.

## ENDNOTES

i The majority of the shares in Telecom Argentina belonged to a joint venture between France Telecom and STET from Italy while Telefónica from Spain controlled Telefónica Argentina. Foreign private investors bought 60% of company shares while the remaining 40% were sold to employees (10%) and the public (30%). Imprecise regulations and delays in the organization of the regulatory agency triggered severe criticism of the process and led to the renegotiation of the concessions for both companies a year after the privatization took place. Therefore, it is considered that the reform was not fully in place until 1991 (Francés, 1993).

ii In its early years of development the Chilean national network consisted of several small local service providers interconnected through the Compañía de Teléfonos de Chile (CTC), owned by the ITT. In 1964 the state stepped into the business opening the Empresa Nacional de Telecomunicaciones (ENTEL-Chile), which coexisted with CTC as a second player in the market. In 1971, during the government of Salvador Allende, CTC was taken over by the government, which attempted to consolidate all companies into a public holding. However, after the 1973 coup, the military government launched a program of liberalization in the industry that allowed the resale of telephone lines and competition in terminal equipment. In 1982 the state began to sell its share in small local companies. And in 1986, it made public offers of its shares in CTC and ENTEL, with Telefónica of Spain emerging as the largest buyer. The changes were accompanied with a reform of the telecom law that freed tariffs in those areas of the country where market conditions were sufficient.

iii TELMEX was developed as a private monopoly until 1971, when the Mexican government bought the majority of the stock owned by ITT and Ericsson. Keeping a minor participation of national private investors, the company grew as a public corporation with subsidiaries in the most diverse businesses (satellite and telegraph, real estate, advertising, and construction) (Francés, 1993). However, a growing technological gap and the deviation of funds to subsidiaries hindered network expansion, and by the end of the 1980s, the underdevelopment of the basic network had become a stumbling block for plans of economic modernization. The government of Carlos Salinas de Gortari justified the privatization of TELMEX as a solution to major bottlenecks in the provision of telecommunication services.

iv The strategy originally set a step-by-step liberalization of the market through the separation of operative and regulatory functions, the corporatization of CANTV, and the liberalization of mobile cellular markets and value-added services. However, the need for multilateral funds to keep macroeconomic balance, and the lack of sources to pay back the considerable foreign debt of the company meant the corporatization strategy was turned into a plan for full privatization in 1990.